# A Distributed Cluster Scheme For Bandwidth Management In Multi-hop MANETs

**Binod Kumar Pattanayak,    Alok Kumar Jagadev,    Manoj Kumar Mishra,    Manoj Ranjan Nayak,**

Siksha 'O' Anusandhan University, Bhubaneswar, India

**Abstract**

Electronic collaboration among devices in a geographically localized environment is made possible with the implementation of IEEE 802.11 based wireless ad hoc networks. Dynamic nature of mobile ad hoc networks(MANETs) may lead to unpredictable intervention of attacks or fault occurrence, which consequently may partition the network, degrade its performance, violate the QoS requirements and most importantly, affect bandwidth allocation to mobile nodes in the network. In this paper, we propose a new distributed cluster scheme for MANETs, especially in harsh environments, based on the concept of survivability to support QoS requirements and to protect bandwidth efficiently. With the incorporation of clustering algorithms in survivability technology, we employ a simple network configuration and expect to reduce occurrences of faults in MANETs. At the same time, we address the scalability problem, which represents a great challenge to network configuration. We do expect a simplification of accessing bandwidth allocation with required QoS support for different applications.

## 1. Introduction

A series of challenges are imposed on support of QoS requirements in MANETs for distributed and real-time multimedia communication, as specified in [1]: a) available bandwidth in a wireless ad hoc network is limited; b) throughput of the network as a whole, is directly affected by degraded radio signals and error prone channels; c)mobility of nodes, limited battery life, service disruption, link failure and security reasons make real-time multimedia communication across a MANET more challenging. Several schemes have been proposed [2, 3] for QoS support in mobile ad hoc networks. However, implementation of proper network topology to meet the QoS requirements, has not yet been clearly determined. On-line QoS requirements in a multi-hop MANET, like end-to-end bandwidth and delay management strictly depends upon network topology. In the absence of a proper network configuration, some nodes might appear to be highly error prone, and as a result, it might be difficult to find a QoS route for the operation of the network.

In the absence of a centralized arbitration mechanism, no fixed topology could be clearly determined for MANETs. Due to mobility of nodes, the network topology in MANETs keep on changing time-to-time. For this reason, MANETs are constrained by limited available bandwidth of links with variable capacity. Hence, it is important to determine a proper topology for MANETs for implementation of high-level routing protocols, which consequently enables it to survive even in harsh environments. Selection of a proper topology for MANETs can be achieved taking into account a set of parameters such as address configuration, transmission power and directions of wireless antennas. Control over network topology allows each wireless node within the network to regulate its parameters (transmitting power, configuration of IP address), so as to achieve a proper topology.

We address in our current research work several aspects related to QoS support to communications in MANETs, and particularly preservation of bandwidth. In addition, we do address a conceptual framework for bandwidth protection, which consequently helps to avoid wasteful allocation of bandwidth or node processing resource in the presence of exception conditions, namely, failure of links, power control and topology portions etc. The basic idea behind this is, to configure a cluster topology, which could be able to meet the QoS requirements, and which could be robust against the unpredictable faults and attacks occurring in MANETs. Thus, a cluster-based network structure possesses the foundations for bandwidth protection in harsh environments. The basic parameters of QoS requirement, addressed by us in this paper, are bandwidth allocation, maximum delay limitations, and most importantly the throughput as a whole.

## 2. Background

### 2.1 QoS support for bandwidth management in MANETs

Support of QoS requirements among the mobile nodes of a MANET strictly depends upon the inherent performance parameters like bandwidth, delay, throughput, errors occurring across links as well as nodes, traffic load within the network, and the algorithms operating at different network layers.

The communications between nodes in a MANET are necessarily affected by obstacles across the network, unpredicted and frequent movements of mobile nodes, externally generated noise and disruptions caused by other communications. It refers to a fact, that quality of links between nodes in a wireless network should necessarily be variable in nature. Preservation of bandwidth for a set of communicating nodes is required, as most applications need to meet specific bandwidth requirements. However, influence of environmental conditions on the wireless communication links makes the task of preservation of bandwidth more complicated.

Adaptive control algorithms, operating at different layers provide a comprehensive approach for preservation of bandwidth in MANETs. Only reservation of bandwidth at a particular layer can not suffice to the QoS requirements of an application in a mobile ad hoc network with some specific traffic characteristics. Desired bandwidth can be obtained, if the communicating nodes adapt traffic to satisfy QoS requirements. The principal goal behind multi-layered approach to bandwidth preservation is to have a control over the bandwidth as perceived by the next higher layer, as the environmental conditions change. Hence adaptation begins at the lower layer and moves to higher layers only when the lower layers are incapable of satisfying the bandwidth requirements at a desired level.

A number of solutions have been proposed for reservation of bandwidth. The existing solutions are mostly categorized as per the layer of the network protocol stack they operate in. A cluster time division multiple access (TDMA) to support real-time traffic in MANETs has been proposed in [5], which effectively utilizes dynamic cluster TDMA for efficient management of limited available resources. A specific cluster can be devised for multiple sessions to share a TDMA, using a code division multiple access (CDMA). An enhanced point coordination function (PCF) mechanism is proposed in [6], that is simpler as compared to the hybrid coordination function (HCF) mechanism.

As proposed by authors in [7], the basic ad hoc on-demand distance vector routing protocols can be enhanced to support QoS requirements, where packets are modified to specify service requirements of nodes forwarding Route Request or Route Reply message, consequently leading to an efficient reservation of bandwidth.

The desired services of applications can be effectively guaranteed by a QoS frame, which incorporates a complete system, where all of its components are bound to cooperate among each other to meet the desired requirements of applications. The conventional models such as IntServ and DiffServ can be effectively implemented in wired networks, with an intention to differentiate services. IntServ is designed on the basis of flow mechanism to provide QoS for bandwidth allocation, and DiffServ is based on classification concept to guarantee QoS. Mobile ad hoc wireless networks can not implement the above two service models for QoS support, due to their inherent characteristics. A flexible QoS model for mobile ad hoc networks (FQMM) have been proposed by Authors in [7], which utilizes per flow granularity of IntServ and aggression of services into classes in Diffserv. This per-flow QoS service adopted in FQMM conveniently overcomes the scalability issues by classifying the low priority services into service classes. In [2], we have proposed a centralized Bandwidth Management Protocol (CBMP) for QoS support to bandwidth allocation in single-hop mobile ad hoc networks. In CBMP, the central component bandwidth manager (BM) is responsible for bandwidth allocation to individual flows in communication with per-flow rate adaptor (RA). However, this scheme does not possess the capability for QoS support to bandwidth allocation in multi-hop mobile ad hoc networks. Currently, we are working over the aspects to bring enhancements into CBMP, so as to make it adaptable to multi-hop MANETs too. In recent times, some QoS signaling schemes for MANETs have already been proposed. The INSIGNIA and the IntServ model share the common feature of soft state bandwidth reservations, which are used to support end-to-end QoS supports for real-time traffic flows. This service level can later be extended to enhanced QoS support, when sufficient resources become available. As per the network dynamics and user-implemented adaptation policies, the framework can scale down, drop or scale up accordingly.

Most of the existing signaling schemes are not designed with a purpose of survivability, and thus, are highly vulnerable to fault occurrences. Unexpected network failures or attacks in harsh or hostile environments may destroy the network topology and disrupt or degrade performance at the application level.

The next section outlines our conceptual frame-work to bandwidth protection in harsh environments.

## 2.2. Problem of bandwidth protection in MANETs

In our current project, we intend to suggest a network architecture, which is adaptable to harsh environments and capable of efficient bandwidth preservation, even in failures or attacks during a communication. The relevance of bandwidth protection in wireless MANETs is demonstrated in Figure.1.

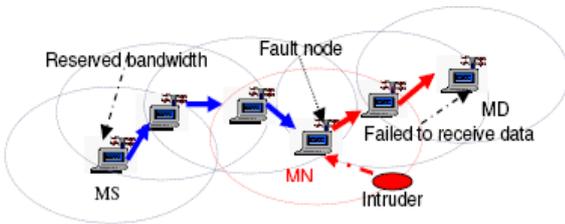

**Figure 1 : Fault occurrence in wireless network**

As shown in Figure 1, the source node MS intends to transmit multimedia data to a destination node MD, for which MS reserves bandwidth along the path. In course of the transmission, at some point of time, an external intruder attacks node MN, along the path of communication between MS and MD. This intrusion causes in link failure, and a portion of network bandwidth is occupied by the intruder. As a result of this failure, the bandwidth along the communication path between nodes MS and MD can not satisfy to the requirements of the application, and consequently destination node MD fails to receive data. Hence, protection of bandwidth during a communication specifically in harsh or hostile environments plays a significant role in successful transmission.

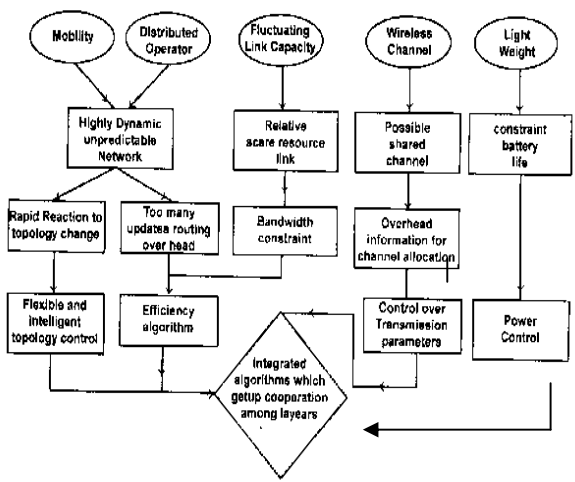

**Figure 2: Cluster-based integrated algorithm**

In our current research work, we concentrate around bandwidth preservation using QoS control at the link level and network layers. Our proposal is oriented around designing a network architecture, which implements distributed, autonomously adaptive algorithms that cooperate to support distributed, real-time multimedia applications and is robust against link failure or attacks in multi-hop mobile ad hoc networks.

The block diagram in Figure.2 demonstrates the outline of the proposed framework. The performance of a mobile ad hoc network is strongly affected by the inherent characteristics of its mobile nodes. Till date, solutions have been proposed for problems of topology control and power control from a standalone point of view. Our proposed framework is about devising an integrated algorithm that incorporates all these aspects. This integrated algorithm can be implemented at a mobile node, specifically in the cluster head of a cluster based algorithm, which follows later.

The network architecture is considered to be steady and adaptive, that is characterized by the use of wireless technology with multiple access, routing protocols as well as other network level protocols. Our proposed architecture is specially tailored at the network layer and MAC layer to guarantee availability even in the presence of failures. Our proposed algorithms are aimed at developing a flexible, robust and efficient network architecture with the implementation of multiple technologies involving the factors, shown in Figure 2. Salient features of these algorithms are a) adaptive control of links, hierarchical organization of a network control structure that reflects location of nodes and network connectivity, b) bandwidth preservation and QoS routing selection, c) autonomous repair of virtual cluster control structure with resource protection.

Our conceptual framework and the proposed clustering algorithm for network configuration are introduced in the following section. The concept of survivability used for network configuration can guarantee communication across the wireless network not disrupting its performance, even in the presence of faults and external attacks.

## 3. Proposed Clustering Scheme

We propose in this section a failure resistant cluster scheme based on the concept of survivability that provides robustness in a distributed and scalable manner.

### 3.1 Conceptual protocol stack

Contribution of our project is to provide a fundament to design of highly mobile wireless network. We do implement a new MANET architecture with autonomous adaptive capabilities. Incorporation of three distinct features in this architecture makes it different from existing schemes: a) distributed network configuration, b) adaptive topology control, c) efficient bandwidth allocation with protection. Figure 3 depicts our conceptual framework with reference to OSI 7-layer protocol stack. We have principally focused in the network layer and MAC layer.

| | | |
|---|---|---|
| **QoS Support** | Multimedia Application | 7 |
| | TCP-UDP (Feedback, non-feedback) | 4-6 |
| | Routing Protocol (Routing with Bandwidth Preservation) | 3 |
| | MAC protocol (Bandwidth Management and Channel Allocation) | 2 |
| | Multi-Hoping | 1 |

**Figure 3** : **Protocol stack for bandwidth preservation**

In our proposed scheme, the network system enables its nodes to organize themselves into a hierarchical control structure. Nodes group themselves autonomously into a cluster around a specific node known as Cluster Head (CH). Nodes are flexible to move from one cluster to another. However, when a node moves to a new cluster, it has to register with CH of the new cluster and deregister from the previous cluster. Depending on the number of nodes in the network, lower level clusters may group themselves to form higher-level clusters.

Routing information in the form of link states, containing information regarding connectivity and available resources belonging to a cluster are propagated across the network by the routing protocols. Each individual CH holds information about connectivity and available resources of all existing clusters within the network. It should be noted that the transmission capability of a cluster is determined by its limited resources, and hence we enable each CH to allocate and manage resources of its own cluster and maintain the resources among other clusters.

Concept of survivability in our scheme is contained in a fact that it not only incorporates robustness against failures resulting from natural fault, but failures induced by malicious adversaries too. The clustering network architecture represents the foundation for efficient bandwidth preservation.

### 3.2 Clustering algorithm

A large spectrum of cluster algorithms for MANETs have been proposed by researchers [10,11,12,13,14]. The cluster algorithms can be classified into two different types, i.e. a) Cluster head (CH) based algorithms; b) distributed and dynamic cluster algorithms. In CH based algorithms, the CH acts as a local control point or local coordinator among the wireless nodes within the cluster. CHs can be further organized into a higher level mesh to establish co-ordination among different clusters. CH-based algorithms carry with themselves the advantages of efficient cluster management and proper load balancing. The distributed and dynamic cluster algorithms deal with adaptive creation of cluster without relying on cluster heads. These algorithms, however, are bound to the limitation that it is difficult to guarantee the number of nodes in a cluster that would be covered. In the course of our research work, we opt to investigate CH-based algorithms with adaptive and agile networking approaches in order to achieve robustness against fault occurrence in mobile ad hoc networks.

It should be noted that the existing cluster-based approaches fail to provide an optimal clustering mechanism, which could consequently provide robustness against fault occurrences. Our proposed scheme incorporates cluster scheme with the concept of survivability which provides the capability to be adaptive to dynamic environments. As per our algorithm, all nodes are allocated to clusters. Cluster head within a specific cluster acts as the access point for that cluster within the wireless network. Procedures of routing incorporate two phases : intra cluster routing and inter cluster routing. Every node in the wireless network has the knowledge about which cluster it belongs to. Each cluster head is availed with the information about the interconnection structure among the clusters (cluster connectivity graph) and the routing paths to neighboring clusters (gateways). In case the two communicating nodes belong to two different clusters, then the routing path between these two clusters is determined first.

Individual clusters and cluster connectivity can be reorganized in a case, if either some critical nodes move significant distances or occurrence of a fault is perceived.

Prior to construction of cluster structure, we start with an assumption that all nodes within a given radius can communicate with each other. Dissemination of small changes in positions of nodes within a cluster, to the cluster head, is avoided.

### 3.3. Fixed grid-cluster construction

Prior to using a network, it should be configured into a set of nodes with network layer connectivity among themselves, supporting addressing, routing and signaling. A MANET can be defined to be a communication graph G = (N,L), where N is a finite set of mobile nodes and L is a set of undirected links among the nodes belonging to N. We presume that all mobile nodes possess an identical radio transmission range, say r .

Clustering of a given network can be achieved by splitting the network into disjoint, adjacent and number of square-routable clusters. As an instance, depending on the size of the network, it can be divided into 4,9,16 clusters etc. Figure 4 shows a clustering scheme with 4 clusters. It can be noticed that each node can register with one CH in its grid.

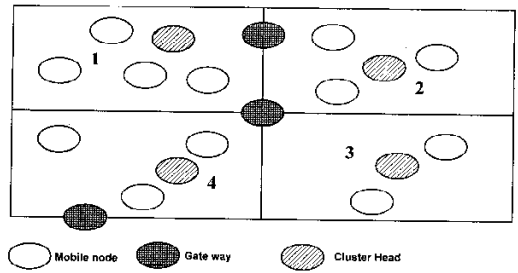

**Figure 4 : Grid cluster structure.**

A unique address (Zone ID) is assigned to each cluster, which represents an ordered pair consisting of the row corresponding to zone and column numbers respectively. It guarantees that the same node can not be a CH in multiple cluster head overlays, and each mobile node is a member of one of the zones. Membership of a node can be determined based on its location in the wireless network. Each node knows which cluster it belongs to, as it has to register with the corresponding cluster head. Each CH knows about how many clusters exist in the network. Nodes existing close to the edge of a cluster can calculate the status of the local gateway and establish gateway connections to the neighboring clusters.

The cluster head (CH) has a temporary role, which changes dynamically as the network topology or the other factors affecting it, keep on changing. A CH is changed on occurrence of any of the following four cases : a) The current CH fails; b) The current CH leaves the current cluster and moves to another; c) The current election period, which is a system design parameter, ends; d) The current CH is replaced by another node, when its load becomes heavier and it needs to move out of the current area. In all above cases, a CH can be replaced by next eligible node in the cluster.

After a cluster is created, a simple selection algorithm runs to select an appropriate mobile node as the CH. The central aspect of selection algorithm is to select the most eligible mobile node in a cluster as cluster head (CH). As elaborated in figure 2, selection of CH depends on a set of parameters like node computation capability, battery power, radio transmission power of the node, node mobility, and location of the node. Selection of CH can be made using the formula, specified in [5]. However, we have modified the formula by introducing a computation factor. We do believe that the node with powerful computational capability deserves to be the CH in a cluster. Location of a node can be determined by control data. Nodes in a cluster exchange control information among each other to select CH. $EF_i$, the eligibility factor of a node i, utilized to serve as a CH at any point of time t, can be obtained as

$$EF_i(t) = a_1 e^{-V_i(t)} + a_2 \left( e^{-I_i(t)} \right) + a_3 B_i^{(t)} + a_3 B_i^{(t)} + a_4 C_i(t) + a_5 (1 - B_i(t)),$$

Where $V_i(t)$ is the average speed of mobile node i at time t, $I_i(t)$ is the frequency node i performing the role of a CH till time t, $B_i(t)$ is the remaining battery power in node i at time t, $C_i(t)$ is the computation status of node i at time t, and $a_1, a_2, a_3, a_4, a_5$ are the weights which reflect the importance of each parameter ( $0 \leq a_k \leq 1$, …..5). The node n with highest eligibility factor $EF_n$ can elect itself as CH in that cluster. The node m with highest eligibility factor $EF_m$ among the rest of the nodes is recorded in CH as the backup CH.

We, hereby, introduce a high level description of the cluster construction algorithm, as follows.

```
Procedure CLUSTER_CONSTRUCTION (num)
num,number of nodes in the mobile wireless network.
begin
    Input num;
    Input desired number of clusters;
    Assign cluster ID to each cluster;
    Mark the cluster number of local node;
    Select CH
    {if((node_grid_id=my_grid_id)
        & (node_coordinations within transmission range
    of local node) & (node_EF>my_EF))
        {Mark the node as potential CH };
        If no other node_EF > potential CH
        {Mark Potential CH as CH };
    If node_EF > ∀ node_i_EF
                    i=1…n
                    node_i ≠ CH
    {Mark node as backup CH}
    if local node is not CH
```

```
            {register local node to CH}};
    Search gateway node
            {if((node_grid ID≠my_grid_id) &
            (node-coordinations within transmission range of
            local node) & (node id≠my_node_id))
            {Mark local node as gateway and
            add the node to gateway list; }
            If a local node is a gateway, add this
            information and propagate it to local CH
            }
end
```

The above algorithm runs at each node at the beginning of network configuration to create clusters and gateway connections between neighboring clusters. This algorithm designates as gateways those mobile nodes, which have connections with nodes in neighboring clusters.

### 3.4 Grid Cluster Management

The routing protocol propagates the cluster information and at the same time guarantees the QoS support to bandwidth protection. A node can automatically recognize a cluster in case it moves to a new cluster. Maintenance of CH is required when gateway status of local node changes. The mobile nodes need to verify its new gateway status in the local cluster and inform it to the CH.

The algorithm for cluster maintenance is summarized below.

```
Procedure CLUSTER_MAINTENANCE
begin
    For each mobile node, after certain node movement or
    time expire;
        Calculate the grid_id of local node.
    If (local node qualities for the gateway)
        {Mark gateway status of local node;
    Check routing table; find gateways in neighboring
    clusters;
    Propagate this information to local CH;
        }
    If (local node was gateway and does not qualify for
    the gateway anymore)
        {Demark local node as gateway;
        Propagate this information to
        local CH;}
end
```

### 4. Analysis

As specified earlier, even though the network may be well configured and can autonomously reconfigure itself with changes in network conditions, the transmission might still be prone to failures due to dynamic nature of wireless mobile networks with unpredictable channel conditions and are quite vulnerable to attacks and failures. Hence, to implement concept of survivability in networks, the underlying protocols need to be sensitive enough to changing conditions in the network. In [15,16], some adaptive applications have been proposed. One of such applications is the topology control via power management.

In our current research work, we have proposed a solution to make use of the concept of survivability, based on active networking technology [17]. This scheme enables MANETs to dynamically select parameters of MAC layer as well as network layer, and to dynamically select suitable protocols based on the requirements of an application as well as the communication environments. In order to meet the bandwidth requirements, the communicating nodes may change from the current MAC layer protocol and routing algorithm to a more capable routing algorithm. In case the environment of a cluster becomes harsher, cognitive networking enables the nodes to learn about the changes in the environment and take required actions for survivability. Robust networking with bandwidth protection strategy can avoid or efficiently deal with exception conditions occurred in the network. Figure 5 depicts our proposed conceptual framework, which integrates MAC layer and network layer together for preservation of bandwidth. The clustering algorithm and fault tolerance algorithm ensure the concept of survivability in the network. The QoS mechanisms ensure the guarantees to applications. Moreover, positioning scheme helps to optimize reservation of bandwidth.

### 5. Conclusion

Our current research work on bandwidth preservation guarantees the application requirements. However, they are still prone to suffer from fault occurrences in harsh environments .

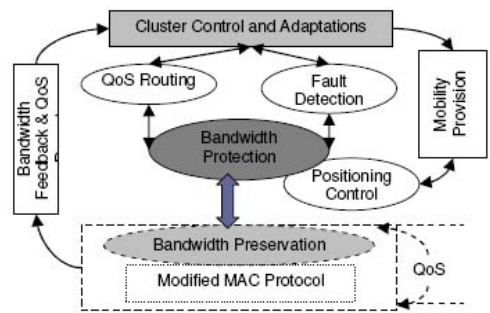

**Figure 5 : Conceptual Architecture Framework For Bandwidth Preservation.**

Our new cluster algorithm based on concept of survivability can provide suitable configurations for survivable multi-hop mobile ad hoc networks. By adopting the concept of cluster creation with concept of survivability we expect to achieve high level of fault tolerance.

## 6. Future Work

Following this conceptual framework we expect to carry out simulations of the proposed clustering scheme using network simulator NS-2. With an intention to enhance survivability of a network, we not only explore the information about network structure and node position, rather we need to explore the information about the type of network. Some geographical routing techniques, especially for sensor networks, have been proposed recently.

Static wireless networks mostly implement topological methods of routing, but some dynamic MANETs can benefit usefully from geographical knowledge about harsh environment. We expect to investigate the ability of nodes to support both of the above mentioned strategies depending on the requirements of applications. In order to make it convenient to create cluster and improve robustness of the wireless network, we plan to develop schemes, which configure IP addresses to nodes automatically.